\documentclass[prl,superscriptaddress,twocolumn]{revtex4-1}
\usepackage[T1]{fontenc}
\usepackage{amsmath}
\usepackage{amssymb}
\usepackage{mathtools}
\usepackage{bm}
\usepackage[normalem]{ulem}
\usepackage{graphicx}
\usepackage{datetime}
\usepackage{color}
\usepackage{times}
\usepackage{braket}
\usepackage{xr}
\usepackage{natbib}
\usepackage{hyperref}
\hypersetup{%
    bookmarksopen=false,
    bookmarksnumbered=true,
    pdfnewwindow=true,
    unicode=false,
    colorlinks=true,%
    citecolor=blue,
    linkcolor=black,
    urlcolor=blue,
    filecolor=blue
    }  
\newcommand{\vb}[1]{\bm{#1}}

\begin{document}
\newcommand{\figdir}{Figures}
\newcommand{\freiburg}{Physikalisches Institut, Albert-Ludwigs-Universit\"{a}t-Freiburg, Hermann-Herder-Stra{\ss}e 3, D-79104, Freiburg, Germany}
\title{Many-body Multifractality throughout Bosonic Superfluid and Mott Insulator Phases}
\author{Jakob Lindinger}
\author{Andreas Buchleitner}
\email[]{a.buchleitner@physik.uni-freiburg.de}
\author{Alberto Rodr\'iguez}
\email[]{Alberto.Rodriguez.Gonzalez@physik.uni-freiburg.de}
\affiliation{\freiburg}

%
\begin{abstract}
We demonstrate many-body multifractality of the Bose-Hubbard Hamiltonian's ground state in Fock space, for arbitrary values of the interparticle interaction. Generalized fractal dimensions unambiguously signal, even for small system sizes, the emergence of a Mott insulator, that cannot, however, be naively identified with a localized phase in Fock space. We show that the scaling of the derivative of any generalized fractal dimension with respect to the interaction strength encodes the critical point of the superfluid to Mott insulator transition, and provides an efficient way to accurately estimate its position. We further establish that the transition can be quantitatively characterized by one single wavefunction amplitude from the exponentially large Fock space. 
\end{abstract}
\maketitle
The properties of a quantum system are crucially determined by the statistical features of its Hamiltonian, as manifestly shown by the applicability of random matrix theory in a variety of scenarios. 
For instance, the system's dynamical behaviour (e.g., the presence of localization, relaxation or long-time equilibration \cite{Borgonovi2016,Madronero2006}), depends decisively on the nature of the eigenenergies and eigenstates, which can exhibit 
high statistical complexity in the form of \emph{multifractality} \cite{Paladin1987,Nakayama2003}.  
Multifractal wavefunctions appear in random matrix models \cite{Mirlin1996,Fyodorov2009,Bogomolny2012c,Kravtsov2015a,Truong2018}, quantum maps \cite{Bogomolny2004,Garcia-Garcia2005,Martin2010,Dubertrand2014,Dubertrand2015a}, and most prominently at the disorder-induced metal-insulator transition \cite{Aoki1983,Aoki1986,Janssen1994,Evers2008}, in the absence (see Refs.~\cite{Rodriguez2011,Ujfalusi2015,Lindinger2017} for recent numerical studies) and in the presence of interactions \cite{Richardella2010,Burmistrov2013,Amini2014,Harashima2014,Burmistrov2015,Carnio2017}. The role of multifractality for this transition in involved geometries \cite{Altshuler2016,Tikhonov2016a,Garcia-Mata2017,Sonner2017,Kravtsov2018} as well as in the many-body localization context \cite{DeLuca2013,Luitz2015,Monthus2016,Pino2017,Serbyn2017} is currently a subject of intense research. 
Interestingly, in the absence of any disorder, multifractality seems to be a generic property of the ground state of many-body spin Hamiltonians \cite{Stephan2009,Stephan2010,Stephan2011,Atas2012,Atas2013}, in which different quantum phases can be identified by corrections to multifractal scaling \cite{Luitz2014,Misguich2017}.
 
\begin{figure}[t]
 \centering
 \includegraphics[width=\columnwidth]{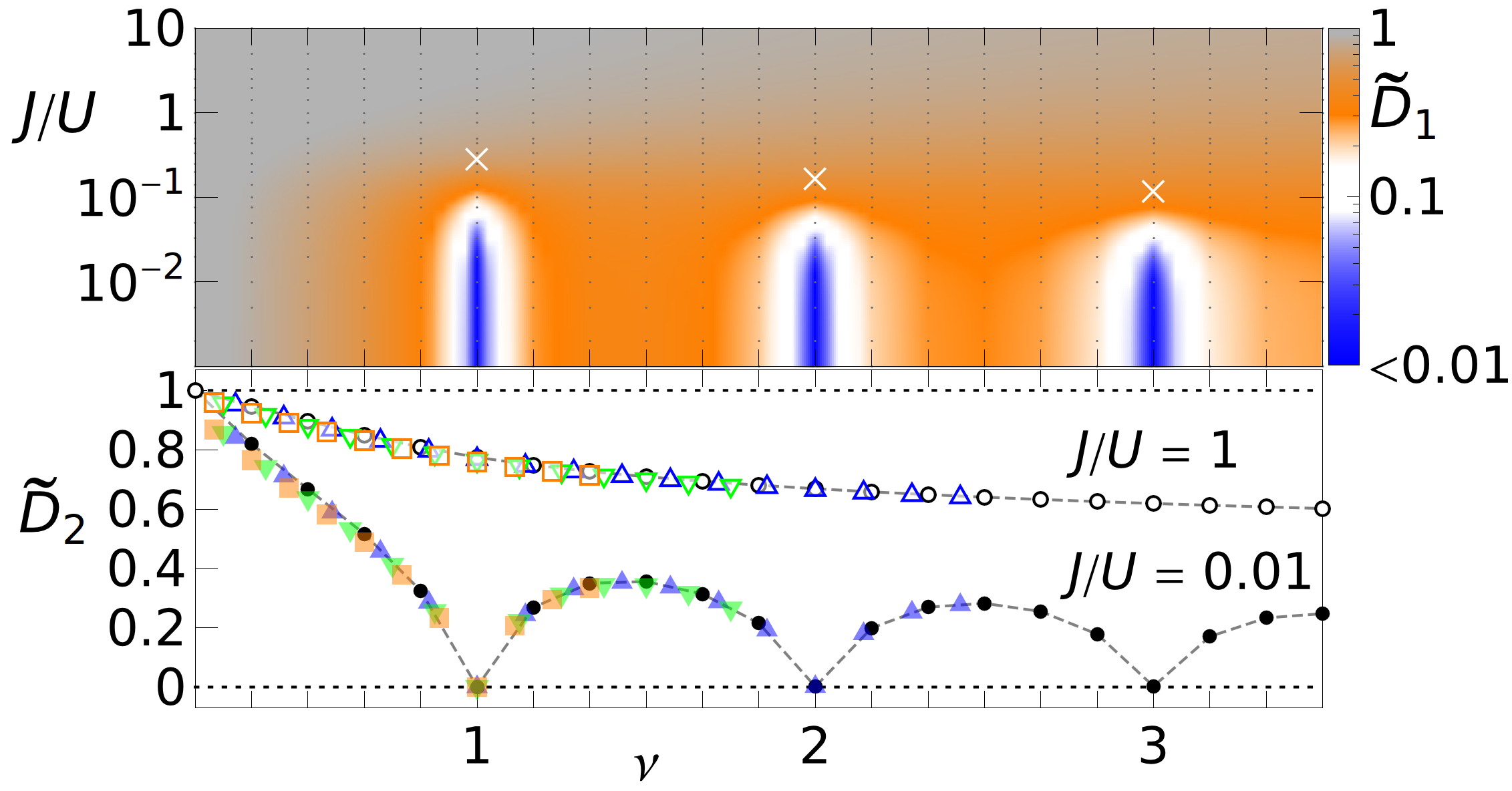}
 \caption{Finite-size fractal dimensions of the BHH ground state versus $J/U$ and filling factor $\nu$
 (abscissa axis in both plots). Upper panel: Density plot of $\widetilde{D}_1$ for $L=6$ after linear interpolation of the numerically calculated points indicated by the black grid. White crosses indicate the position of the SF to MI transition \cite{Carrasquilla2013,Ejima2011}. Lower panel: $\widetilde{D}_2$ for $J/U=1$ (open symbols), $J/U=10^{-2}$ (filled symbols) and $L=6$ (black), $7$ (blue), $8$ (green), $9$ (orange).}
 \label{fig:DqDenplot}
\end{figure}

In this work, we demonstrate that the statistical complexity of many-body states in `clean' (not disordered) bosonic systems can be described in terms of multifractality. Such characterization can not only provide an unambiguous identification of localized, extended and ergodic wavefunctions, but also exposes how the presence of different macroscopic properties (phases) of the system is rooted in the Hilbert-space structure of quantum states. Here, we exemplify the potential of such analysis by showing that the multifractal properties of the Bose-Hubbard Hamiltonian (BHH) ground state in the Fock basis carry a  distinctive signature of the transition from superfluid (SF) to Mott insulator (MI), as shown in Fig.~\ref{fig:DqDenplot}. 
This novel approach reveals that the transition is fully encoded in the behavior of one single wavefunction amplitude in Fock space, and it further provides an efficient way to locate the critical point.

Let us consider the expansion of a quantum state 
in an orthonormal basis of the underlying Hilbert space of size $\mathcal{N}$, $\ket{\Psi}=\sum_{j=1}^\mathcal{N} \psi_j \ket{j}$,
and define the $q$-moments of the distribution of intensities as 
$ R_q=\sum_{j=1}^\mathcal{N} |\psi_j|^{2q}$, for $q\in\mathbb{R}^+$.
The scaling of $R_q$ with $\mathcal{N}$ reveals the asymptotic statistics (for large $\mathcal{N}$) of the participation of the basis elements $\ket{j}$ in the state $\ket{\Psi}$. Such scaling is generically expected to be 
of the form 
 $R_q\sim \mathcal{N}^{-(q-1)D_q}$,
where $0\leqslant D_q\leqslant 1$ are the \emph{generalized fractal dimensions} (GFD). An ergodic extended state in the considered basis, defined by $|\psi_j|^2\sim \mathcal{N}^{-1}$ as $\mathcal{N}\rightarrow\infty$, has $D_q=1$ for all $q$. On the other hand, if for any $q>1$ saturation of $R_q$ with $\mathcal{N}$ is observed we speak of localized states, for which, consequently, $D_{q>1}=0$ \cite{Note1}.
If $q$-dependent values 
$0<D_q<1$ occur, the state is \emph{multifractal} in the $\ket{j}$ basis \cite{Paladin1987,Janssen1994,Evers2008}. 
The parameter $q$ controls which wavefunction intensity (roughly, which value of $-\log_{\mathcal{N}}|\psi_j|^2$) dominates $R_q$ for large $\mathcal{N}$. 
Thus, different dimensions $D_q$ ensue if each set of points in the wavefunction with a certain intensity scales differently when enlarging the Hilbert space accessible to $\ket{\Psi}$, i.e., 
if the volume of each intensity set scales as a power law of $\mathcal{N}$ with its own (in general non-integer) exponent. In such case, each intensity set is a fractal, and their superposition makes up the multifractal state. 

In order to characterize multifractality, it is useful to define $\mathcal{N}$-dependent dimensions $\widetilde{D}_q$, 
\begin{equation}
 \widetilde{D}_q\equiv \frac{1}{1-q} \log_\mathcal{N} R_q,
\end{equation}
that for increasing $\mathcal{N}$ converge to the GFD, $D_q=\lim_{\mathcal{N\rightarrow\infty}} \widetilde{D}_q$.
Note that $\widetilde{D}_q=S_q/\ln\mathcal{N}$, where $S_q$ is the Shannon-R\'enyi entropy of degree $q$ of the $|\psi_j|^2$ distribution. 
Among the GFD, we single out the 
cases
$q=1,2$ and $\infty$. The exponent $D_1$ is known as the \emph{information dimension} since it determines the scaling of the Shannon information entropy, $-\sum_j|\psi_j|^2\ln|\psi_j|^2 \sim D_1 \ln \mathcal{N}$. The dimension $D_2$ controls the growth of the participation ratio, $R_2^{-1}$, which is regarded as a measure of the `volume' of the state: Finite and $\mathcal{N}$-independent for localized wavefunctions but unbounded for extended states. For a multifractal state $R_2^{-1}\sim \mathcal{N}^{D_2}$, i.e., its `volume' diverges with $\mathcal{N}$ but it occupies a vanishing fraction of the total Hilbert space. Multifractal wavefunctions are therefore an example of non-ergodic extended states. 
For $q=\infty$ the moments $R_q$ are determined by the maximum value of the intensities, $p_\text{max}\equiv\max_j |\psi_j|^2$, and  $\widetilde{D}_\infty=-\log_\mathcal{N} p_\text{max}$. The GFD as well as their finite-size counterparts are always monotonously decreasing functions of $q$ \cite{Hentschel1983}. Hence, the minimum GFD is $D_\infty$ ($\widetilde{D}_\infty$ for fixed $\mathcal{N}$).

We apply this formalism to analyze the statistical properties of the ground state of the BHH in one dimension (1-D) \cite{Lewenstein2007,Cazalilla2011,Krutitsky2016}, which in terms of bosonic annihilation and creation operators, $b_k$, $b_k^\dagger$, $n_k\equiv b_k^\dagger b_k$, reads 
\begin{equation}
H= -\eta\sum_k (b^\dagger_k b_{k+1}+b^\dagger_{k+1} b_k) + \frac{1}{2}\sum_k n_k(n_k-1),
 \label{eq:BHH}
\end{equation}
where $\eta\equiv J/U$ is the ratio of hopping to interaction strength ($U>0$).
Our system includes $N$ bosons in $L$ lattice sites with periodic boundary conditions (PBC). In the thermodynamic limit ($N,L\rightarrow\infty$), at fixed integer filling factor $\nu\equiv N/L$, the ground state of $H$ undergoes a Berezinskii-Kosterlitz-Thouless (BKT) phase transition as a function 
$\eta$, between a MI and a SF state \cite{Fisher1989,Greiner2002,Bakr2010}.  In 1-D, the position of the critical point for $\nu=1$ has been estimated to be $\eta_c\simeq 0.3$, both theoretically (see Refs.~\cite{Rachel2012,Gerster2016,Krutitsky2016} and references therein) and experimentally \cite{Boeris2016}. 

A convenient basis of the Hilbert space of $H$, of size $\mathcal{N}=\begin{psmallmatrix} N+L-1 \\ N\end{psmallmatrix}$,  is given by the Fock states of the on-site density operators, 
$\ket{\boldsymbol{n}}\equiv\ket{n_1,n_2,\ldots,n_L}$, where $||\bm{n}||_1=N$. Hence, the ground state of the system can be expanded as  
$ \ket{\Psi(\eta)}=\sum_{\boldsymbol{n}} \psi_{\boldsymbol{n}}(\eta) \ket{\boldsymbol{n}}$.
For integer $\nu$ and $\eta\rightarrow0$, the ground state is given by one element of the Fock basis,  
\begin{equation}
 \ket{\Psi(0)}=\ket{\nu,\nu,\ldots,\nu}\equiv\ket{\bm{\nu}}.
 \label{eq:Psi2Uinf}
\end{equation}
Conversely, 
in the non-interacting limit ($\eta\rightarrow\infty$) the intensities of $\ket{\Psi}$ converge to
\begin{equation}
 |\psi_{\bm{n}}(\infty)|^2=\frac{N!}{L^N n_1!n_2!\ldots n_L!},
 \label{eq:Psi2U0}
\end{equation}
and the full Fock basis participates in the state. 
The extremely localized nature of the ground state for $\eta=0$ leads to $\widetilde{D}_{q>0}=D_{q>0}=0$. 
For $\eta\rightarrow\infty$, 
the GFD can also be analytically obtained, and 
have non-trivial $q$-dependent values, e.g., for $\nu=1$, $D_1=0.941$, $D_2=0.907$, $D_\infty=(2\ln2)^{-1}=0.721$, i.e., the ground state exhibits multifractality in the Fock basis 
\cite{Note2,Lindinger2017c}.

\begin{figure}[t]
 \centering
 \includegraphics[width=.9\columnwidth]{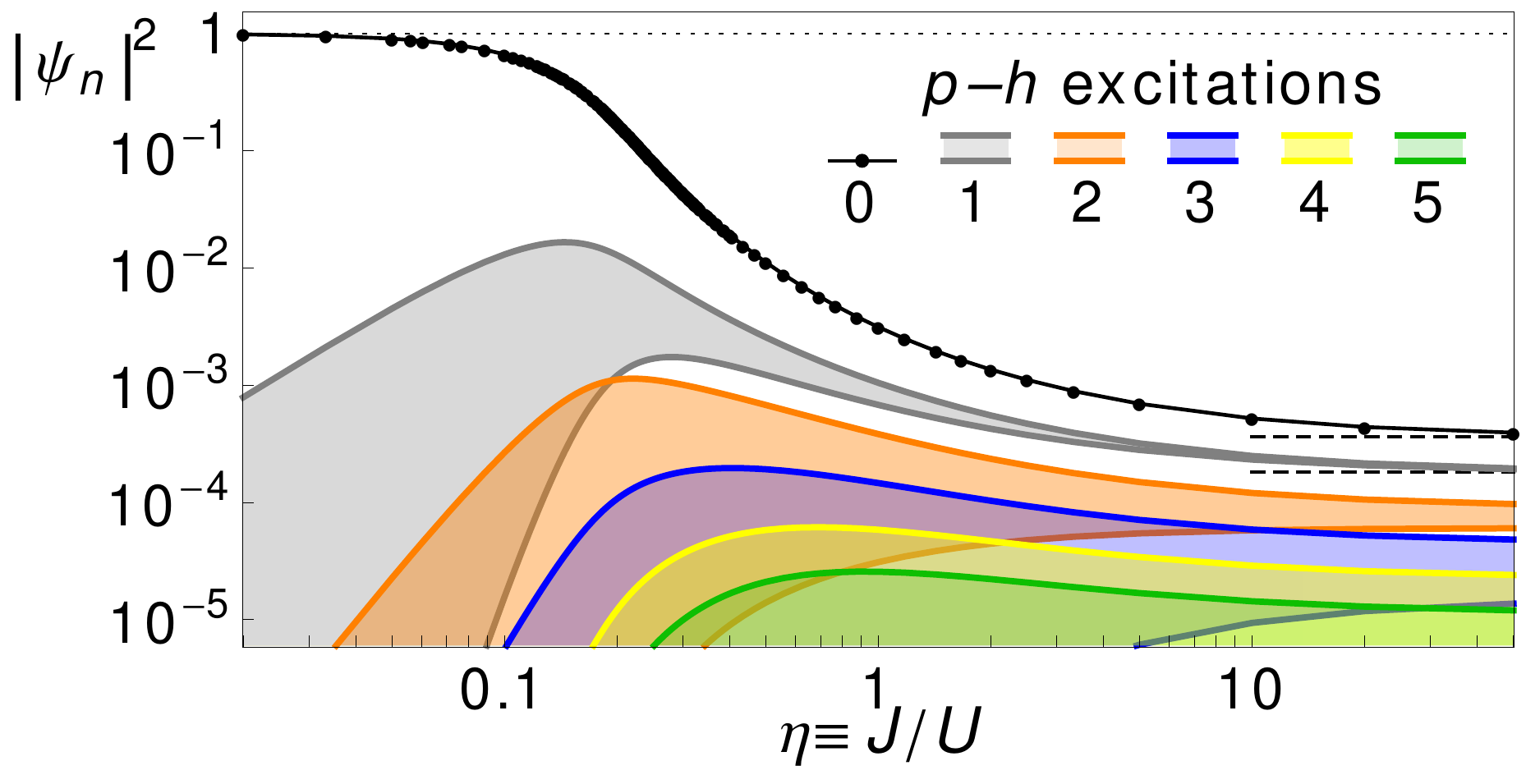}
 \caption{Intensities $|\psi_{\vb{n}}|^2$ in Fock basis of the BHH ground state versus $\eta$ for $L=10$, $\nu=1$. Solid lines highlight the maximum and minimum intensities on a Fock state with a certain number of particle-hole ($p$-$h$) excitations on top of the homogeneous state $\ket{\bm{\nu}}$. The values of $\eta$ considered are highlighted by symbols only for the maximum intensity. Dashed lines indicate the intensity value of the first two $p$-$h$ manifolds for $\eta=\infty$ [see Eq.~\eqref{eq:Psi2U0}].} 
 \label{fig:Psi2vU}
\end{figure}

How do the GFD evolve with $\eta$ between these two limits, and does this evolution expose the MI-SF transition? 
For such intermediate values of $\eta$, the multifractal analysis must be performed numerically:
We combine exact diagonalization (for systems $L\leqslant 10$) with a recently proposed technique \cite{Luitz2014,Luitz2014d,Luitz2014f,Luitz2014g} based on quantum Monte Carlo (QMC) to estimate the moments $R_q$ for larger systems efficiently 
\cite{Note3}.

Remarkably, the analysis of the finite-size dimensions $\widetilde{D}_q$ for different $\eta$ and varying filling factor reveals a distinct and unambiguous signal 
of the emergence of a MI state, as demonstrated in Fig.~\ref{fig:DqDenplot}.
Whereas for weak interaction the finite-size GFD change monotonously with $\nu$, they register a pronounced drop towards zero at integer densities in a range of $\eta$ that clearly correlates with the MI phase. 
For integer density, all $\widetilde{D}_q$ vanish asymptotically as $\eta\rightarrow0$. For non-integer filling, however, all $\widetilde{D}_q$ remain finite as $\eta\rightarrow0$, signaling the persistence of a SF phase for any value of the interaction. 

Let us further note that, for integer $\nu$, according to Eqs.~\eqref{eq:Psi2Uinf} and \eqref{eq:Psi2U0}, the maximum intensity of $\ket{\Psi(\eta)}$ in the Fock basis occurs for the homogeneous state $\ket{\bm{\nu}}$ in both limits $\eta=0$ and $\eta=\infty$. 
The hopping and interaction terms of $H$ minimize independently 
the energy by maximizing the amplitude on $\ket{\bm{\nu}}$. This property persists for any value of $\eta$ and $L$ (when using PBC) 
as illustrated in Fig.~\ref{fig:Psi2vU}. 
This makes the dimension $\widetilde{D}_\infty$ particularly accessible, since it will be entirely determined by the probability $\left|\braket{\bm{\nu}|\Psi(\eta)}\right|^2$, which can be straightforwardly estimated using QMC.  

\begin{figure}[t]
 \includegraphics[width=.95\columnwidth]{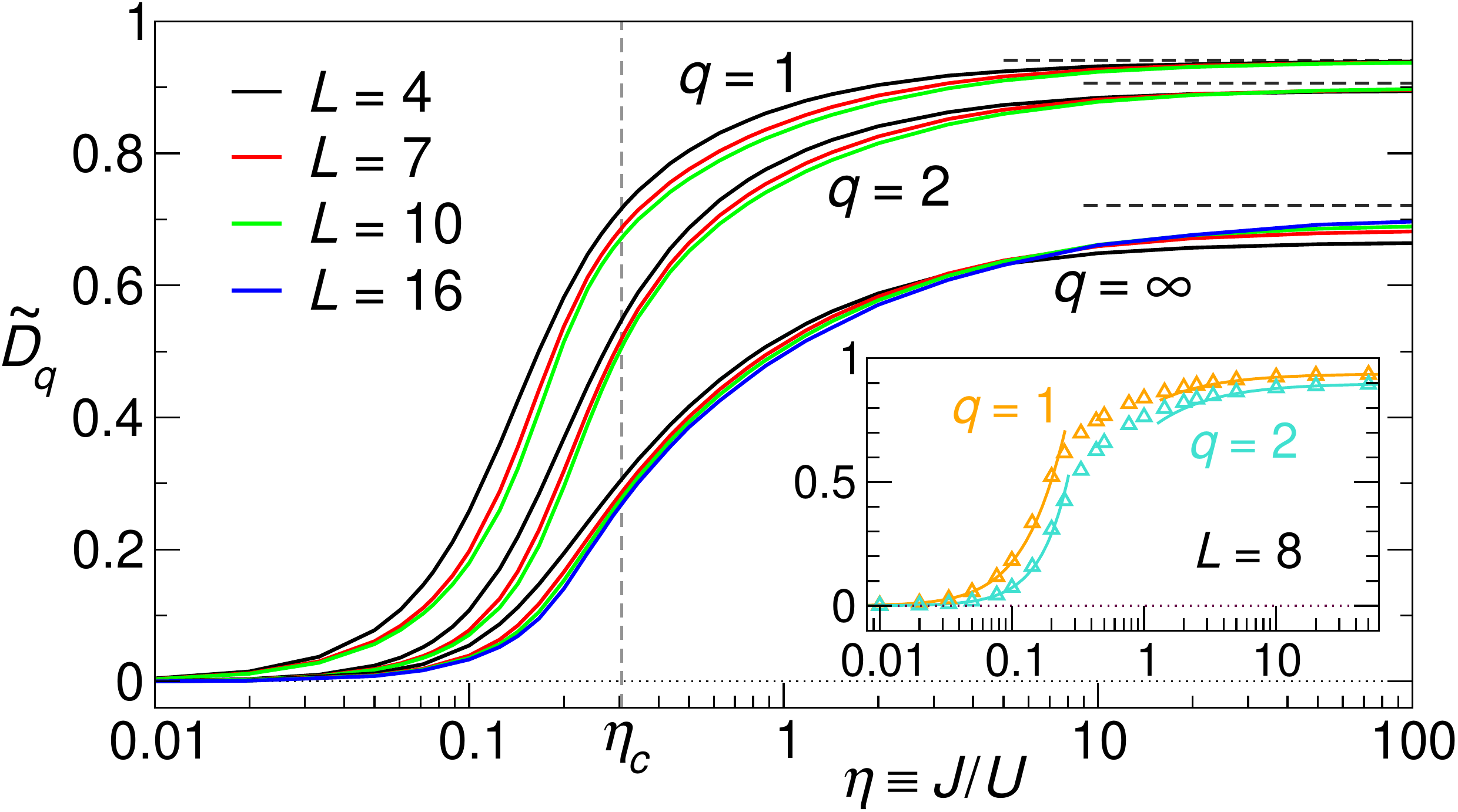}
 \caption{Finite-size GFD  $\widetilde{D}_q$ ($q=1,2,\infty$) of the BHH ground state versus $\eta$ for $\nu=1$. Solid lines in main panel are numerical results ($L=16$ only for $q=\infty$). Horizontal dashed lines mark the $D_q$ values for $\eta=\infty$. The inset shows numerical (symbols) and analytical results from perturbation theory (solid lines) for $L=8$.}
 \label{fig:DqOverview}
\end{figure}
An overview of the $\eta$-dependence  of $\widetilde{D}_q$, for $q=1,2,\infty$, $\nu=1$ and different $L$ is shown in Fig.~\ref{fig:DqOverview}. 
The value of the finite-size GFD is strongly suppressed for small $\eta$ and rises quickly as the vicinity of the critical value $\eta_c$ is approached. The SF phase thus correlates with higher values of $\widetilde{D}_q$, indicating a faster growth of the ground state's `volume' in Fock space as $L\rightarrow\infty$. 
Although convergence towards the thermodynamic limit $D_q$ is rather slow, the data strongly suggest that multifractality exists for any $\eta$. We emphasize that for $\eta\ll 1$ and $\eta\gg 1$ the GFD for small $L$ are very well described by perturbation theory \cite{Lindinger2017c}, as shown in the inset of Fig.~\ref{fig:DqOverview}. 

From the behavior observed in Fig.~\ref{fig:DqOverview}, it is rather appealing to think that $\widetilde{D}_q$ may vanish in the thermodynamic limit for $\eta\leqslant\eta_c$. In such a case, the MI phase would have a simple interpretation as a localized phase in Fock space ---which can be viewed as an intricate lattice, whose nodes, i.e., the states $\ket{\bm{n}}$, have different energies and are connected by the hopping term of Hamiltonian \eqref{eq:BHH}. Nevertheless, as $\mathcal{N}\rightarrow\infty$, the coordination number of the Fock lattice diverges linearly with $L$, and therefore, naively, the existence of localization in the thermodynamic limit appears unlikely. In order to ascertain the presence or absence of localization in the MI phase, a proper $\mathcal{N}\rightarrow\infty$ extrapolation is required, for which knowledge of the expected finite-size corrections is essential. The analytical calculation of the GFD in the non-interacting limit provides access to the leading finite-$L$ corrections, whose form is essentially determined by the dependence of $\mathcal{N}$ on $L$. Using insights from perturbation theory and the analysis of plausible asymptotic behaviors of $R_q$, we find that the dominant finite-size corrections in the scaling of the GFD for any $\eta$ are 
\begin{equation}
 \widetilde{D}_q=D_q+\alpha\frac{\ln L}{L} +\beta\frac{1}{L}+\gamma\frac{\ln^2 L}{L^2}+\mathcal{O}(L^{-2}\ln L),
 \label{eq:DqScaling}
\end{equation}
with $\eta$- and $\nu$-dependent coefficients $\alpha$, $\beta$, $\gamma$.

\begin{figure}[t]
 \includegraphics[width=.95\columnwidth]{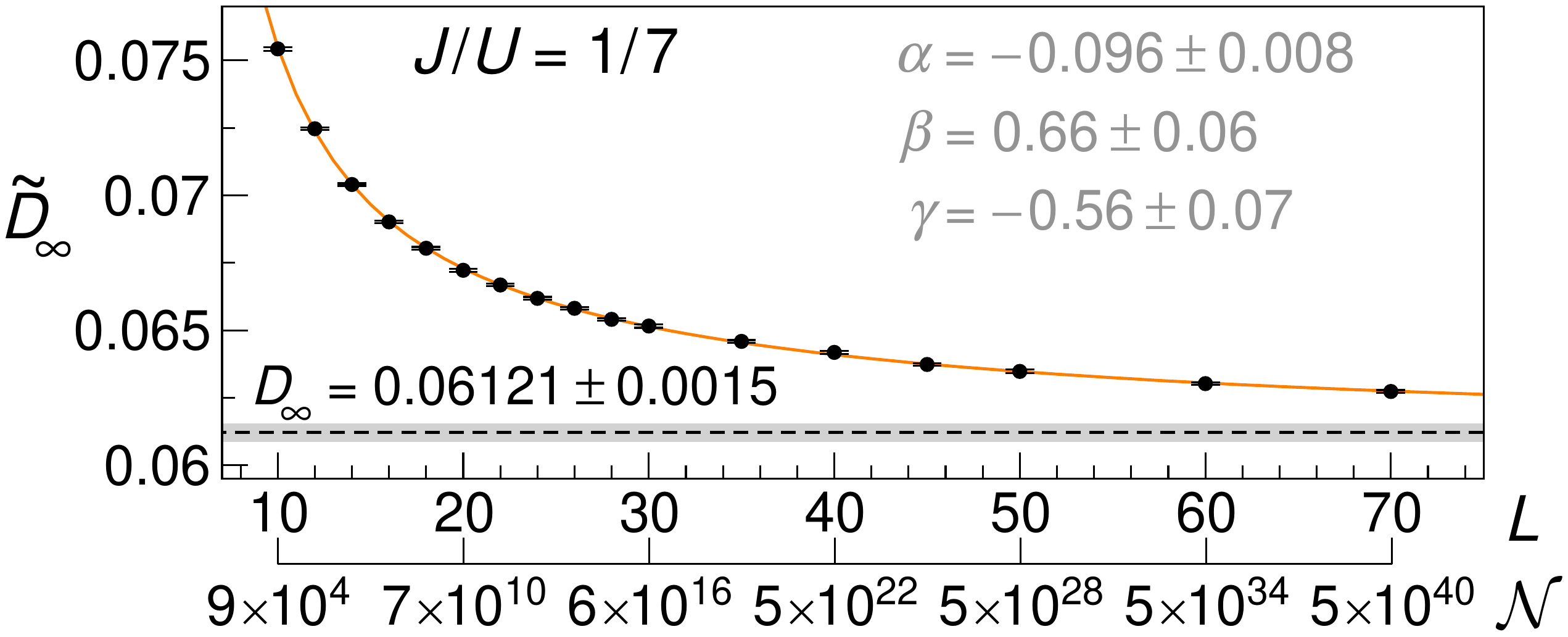}
 \caption{Extrapolation of $\widetilde{D}_\infty$ as $L\rightarrow\infty$ for $\eta=1/7$, $\nu=1$. Symbols are numerical data. The solid line is the best fit to Eq.~\eqref{eq:DqScaling} (chi-square $\simeq 10$ with $13$ degrees of freedom). The horizontal dashed line and the shaded area mark, respectively, the $D_\infty$ value and its 95\% confidence interval. The secondary abscissa axis indicates the size of Fock space for each $L$.}
 \label{fig:DinfLtoinf}
\end{figure}

We analyzed the minimum dimension $\widetilde{D}_\infty$ for system sizes up to $L=70$ at unit filling for $\eta=1/7<\eta_c$. The numerical data is perfectly described by the first four terms in Eq.~\eqref{eq:DqScaling}, as shown in Fig.~\ref{fig:DinfLtoinf}. Indeed, only if the four terms are present can a reliable fit be obtained. The resulting $D_\infty$ is distinctively non-vanishing (consequently $D_q>0$ for all $q$), and hence Fock-space localization in the MI phase is \emph{ruled out}. 

We conclude that there is no fingerprint of the transition in the raw values of the GFD: In the thermodynamic limit, the dependence of $D_q$ with $\eta$ will exhibit an overall behavior similar to that observed in Fig.~\ref{fig:DqOverview} for finite Fock spaces. Yet the evolution of the GFD might still encode the transition. The MI-SF crossover for finite $L$ has recently been inspected from another perspective: In 2-D via the $\eta$-derivatives of the expectation value of simple observables \cite{Lacki2016}, and in 1-D using the fidelity susceptibility \cite{Buonsante2007,Carrasquilla2013,Lacki2014}. 
The common underlying idea to these approaches is to use the $\eta$-sensitivity of the ground state as a figure of merit. In our formalism, the $\eta$-dependence of the dimensions $\widetilde{D}_q$ exposes manifestly the variation in the structure of the ground state in Fock space, and, consequently, we find that the rate of change of the GFD with $\eta$ reveals the critical point. 
 
\begin{figure}
 \includegraphics[width=\columnwidth]{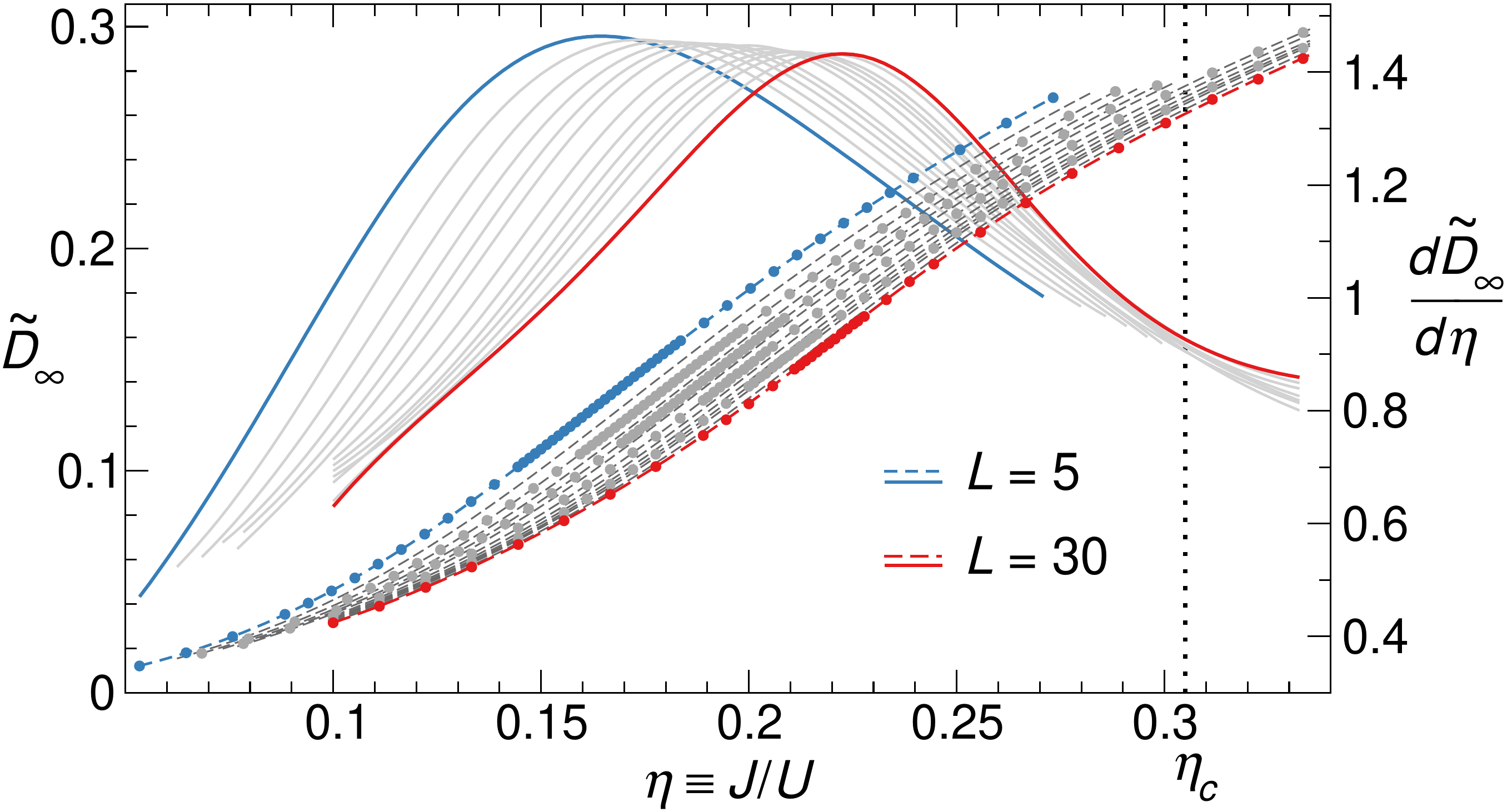}
 \caption{Dimension $\widetilde{D}_\infty$ (left vertical axis) and its derivative (right vertical axis) versus $\eta$ for the BHH ground state and $L=\{5\text{--}10,12,14,16,18,20,25,30\}$, $\nu=1$. Symbols indicate numerical data (errors within symbol size), dashed lines correspond to the best Pad\'e fits and solid lines are their respective derivatives. For clarity, symbols are shown only for $L=\{5,7,9,12,18,30\}$.}
 \label{fig:PadefitsDinf}
\end{figure}
In Fig.~\ref{fig:PadefitsDinf}, we show $\widetilde{D}_\infty$ as a function of $\eta$ and its corresponding derivative $\widetilde{D}'_\infty(\eta)\equiv d \widetilde{D}_\infty/d\eta$ for $L\leqslant 30$. When approaching the transition from the MI side, the derivative develops a distinct single maximum at a certain value $\eta_*(L)$ that shifts towards $\eta_c$ for increasing $L$. In order to locate reliably the position of the maximum, we first find the best fit of the numerical $\widetilde{D}_\infty$ data to a Pad\'e approximant, which is then differentiated.  
The analysis of $\widetilde{D}_2$ reveals the same behavior \cite{Note4}.
The scaling expected for the position of the maximum of the derivative follows from the assumption that at $\eta_*(L)$ the correlation length $\xi$ \cite{Lewenstein2007} (ruling the spatial decay of the single-particle density matrix elements, $\langle b^\dagger_kb_{k+r}\rangle\sim e^{-r/\xi}$) is proportional to the system size. While $\xi$ is finite and $L$-independent in the MI phase, it diverges at the transition and within the SF phase \cite{Note5}. We expect that the steepest change of each GFD with increasing $\eta$ correlates with the region where $\xi\sim L$, i.e., it signals the crossover for a finite system.  
For $\eta<\eta_c$ the correlation length exhibits the exponential dependence $\xi\sim \exp(b/\sqrt{\eta_c-\eta})$ with $b>0$. Hence, it ensues 
\begin{equation}
 \eta_*(L)=\eta_c-\frac{b^2}{\ln^2(L/|\ell_q|)},
 \label{eq:ScalingMaxDer}
\end{equation}
for suitable parameters $\eta_c$, $b$ (which are $q$ independent) and $\ell_q$. 
Note that this same scaling holds for the position of the maximum of the fidelity susceptibility at a BKT transition \cite{Sun2015a}.

\begin{figure}
 \includegraphics[width=.9\columnwidth]{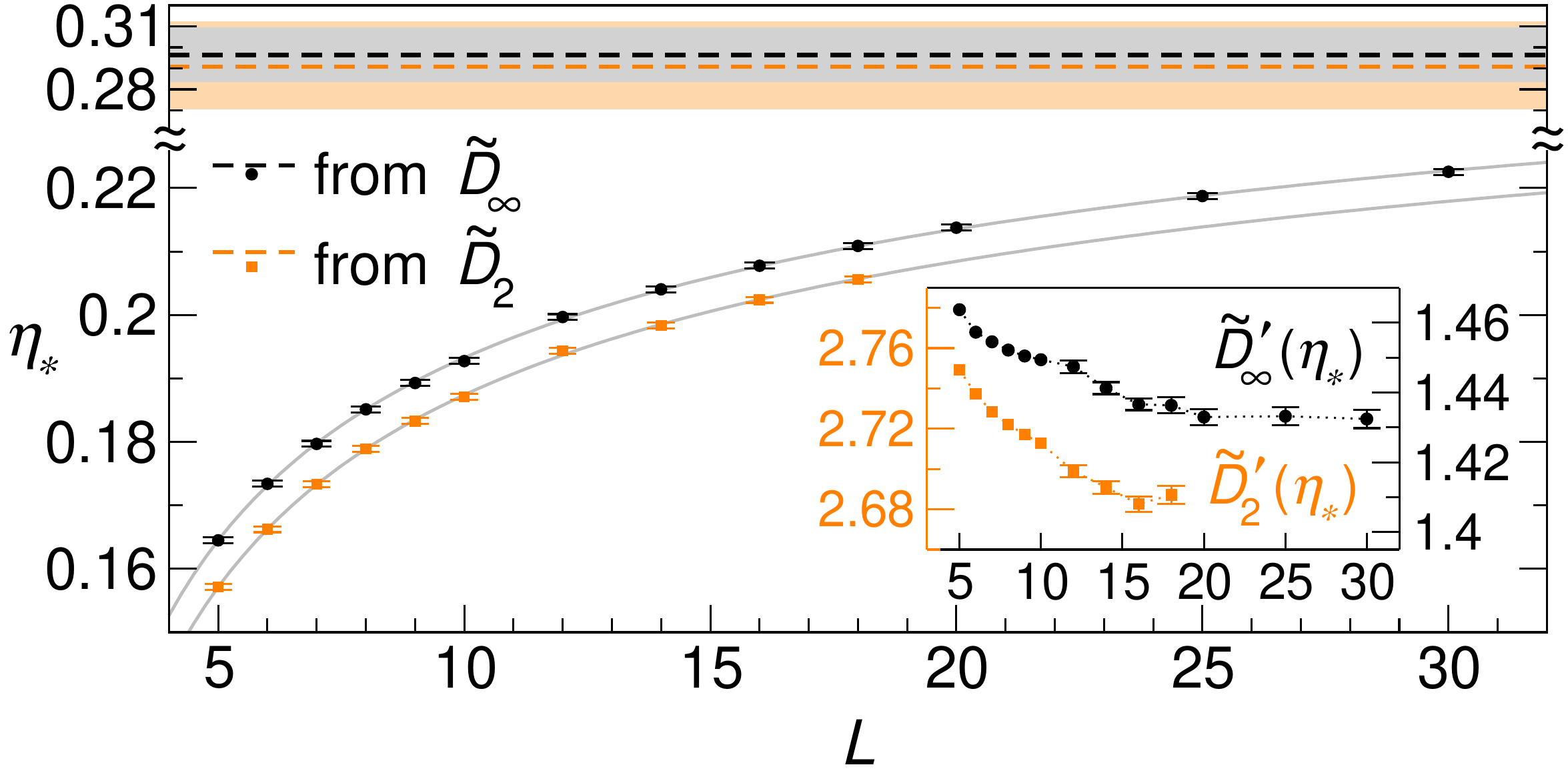}
 \caption{Position $\eta_*(L)$ of the maximum of $\widetilde{D}'_\infty(\eta)$ and $\widetilde{D}'_2(\eta)$ for $\nu=1$. Solid lines are best fits to Eq.~\eqref{eq:ScalingMaxDer}, with $b=1.92\pm0.18\;(1.85\pm0.27)$, $\ell_q=0.025\pm0.009\;(0.032\pm0.018)$ for $\widetilde{D}_\infty$ ($\widetilde{D}_2$) data. Dashed lines and shaded regions mark, respectively, the estimated $\eta_c$ and its 95\% confidence interval: $\eta_c \in[0.284,0.308]$ from $\widetilde{D}_\infty$, and 
  $\eta_c\in[0.270, 0.312]$ from $\widetilde{D}_2$. The inset shows the maximum value of the derivatives versus $L$.} 
 \label{fig:MaxDer}
\end{figure}
The scaling analysis of $\eta_*(L)$ is presented in Fig.~\ref{fig:MaxDer} for $\widetilde{D}_\infty$ ($L\leqslant 30$) and for $\widetilde{D}_2$ ($L\leqslant 18$). The data is described reliably by Eq.~\eqref{eq:ScalingMaxDer}, which yields the following estimates for the critical point at unit filling:
$\eta_c=0.296\pm0.006$ 
from 
$\widetilde{D}_\infty$, and 
$\eta_c=0.291\pm0.011$ 
from 
$\widetilde{D}_2$. Both values are compatible with each other and in perfect agreement with previous estimates. 
The maximum value of the derivatives seems to be finite for $L\rightarrow\infty$ (inset of Fig.~\ref{fig:MaxDer}), which has similarly been observed for the fidelity susceptibility at a BKT transition \cite{Sun2015a}.
The scaling of the 
GFD derivatives provides a very good level of accuracy in the location of the critical point already from the analysis of modest system sizes. Furthermore,  
$\widetilde{D}_\infty$ is simply the value of the intensity $\left|\braket{\bm{\nu}|\Psi(\eta)}\right|^2$ in log scale, hence, we have demonstrated that the MI-SF transition can be characterized by monitoring only one wavefunction amplitude in the exponentially large Fock space \cite{Note6}. 

We have provided evidence of the significance of many-body multifractality in Fock space for bosonic systems. In particular, we have shown that the superfluid to Mott insulator transition in the Bose-Hubbard Hamiltonian (BHH) can be understood in terms of the rate of change of the generalized fractal dimensions (GFD) with the interaction strength. Such novel perspective provides an efficient method to locate accurately the critical point using moderate system sizes. Remarkably, it furthermore reveals that the transition at integer densities can be analyzed from the examination of only one privileged (maximum) wavefunction intensity. This observation 
opens a promising path for further theoretical and experimental studies of the BHH. 
We also note that the $\eta$-dependence of the information entropy has been used to characterize the statistical nature of the BHH eigenstates \cite{Kollath2010} and to identify chaotic behaviour \cite{Kolovsky2004}.
Additionally, first results indicate that multifractality extends to the excited states, whose GFD also carry apparently a fingerprint of the transition. Whereas the rather expected absence of localization in Fock space has been confirmed, it remains to be seen whether (many-body) Fock-localized phases exist for the disordered BHH. 

\begin{acknowledgments}
We are grateful to V.~G.~Rousseau for providing the QMC code and to L.~de Forges de Parny for helpful discussions. Furthermore, A.~R.~thanks D.~Luitz for useful discussions. 
The authors acknowledge support by the state of Baden-W\"urttemberg through bwHPC and the German Research Foundation (DFG) through grant no 402552777.
\end{acknowledgments}
\bibliographystyle{apsrev4-1}
%
\newpage 
\newcommand{\TITLE}{Many-body Multifractality throughout Bosonic Superfluid and Mott Insulator Phases}
\onecolumngrid
\begin{center}\large\bfseries Supplemental Material \\[2mm] \TITLE \end{center}
\vskip 4mm
\twocolumngrid
\renewcommand\theequation{S\arabic{equation}}
\renewcommand\thefigure{S\arabic{figure}}
\section{Numerical estimation and analysis of \boldmath$\widetilde{D}_q$}

The finite-size dimensions $\widetilde{D}_q$ can be straightforwardly estimated if the coefficients $\psi_{\bm{n}}(\eta)$ of the ground state are known. Using exact diagonalization (ED) it is possible to obtain the ground state in the Fock basis for systems $L\lesssim 12$ at unit filling, but the exponential growth of the size of the Fock space impedes the use of this approach for larger system sizes. 
As shown in Eq.~\eqref{eq:Psi2Uinf}, the full basis participates in the state for $\eta=\infty$, while for $\eta=0$ we have $\ket{\Psi(0)}=\ket{\nu,\ldots,\nu}\equiv\ket{\bm{\nu}}$. For integer density, one can check that the maximum wavefunction intensity corresponds to the Fock state $\ket{\bm{\nu}}$ for any value of $\eta$ and $L$, as long as periodic boundary conditions (PBC) are applied. The second largest intensity is always linked to a Fock state with the lowest energy particle-hole ($p$-$h$) excitation on top of $\ket{\bm{\nu}}$, i.e.~to $\ket{\bm{\nu}\pm1}\equiv\ket{\nu,\ldots,\nu-1,\nu+1,\ldots,\nu}$ (or the corresponding normalized translationally invariant superposition thereof). 
Hence,
\begin{equation}
 \ket{\Psi(\eta)}=\psi_{\bm{\nu}}(\eta)\ket{\bm{\nu}} + \psi_{\bm{\nu}\pm1}(\eta)\ket{\bm{\nu}\pm1}+\ldots,
\end{equation}
where the remaining terms involve coefficients $\psi_{\bm{n}}(\eta)$ with smaller intensities. This is demonstrated in Fig.~\ref{fig:Psi2vU}.

\begin{figure}[b]
 \includegraphics[width=.95\columnwidth]{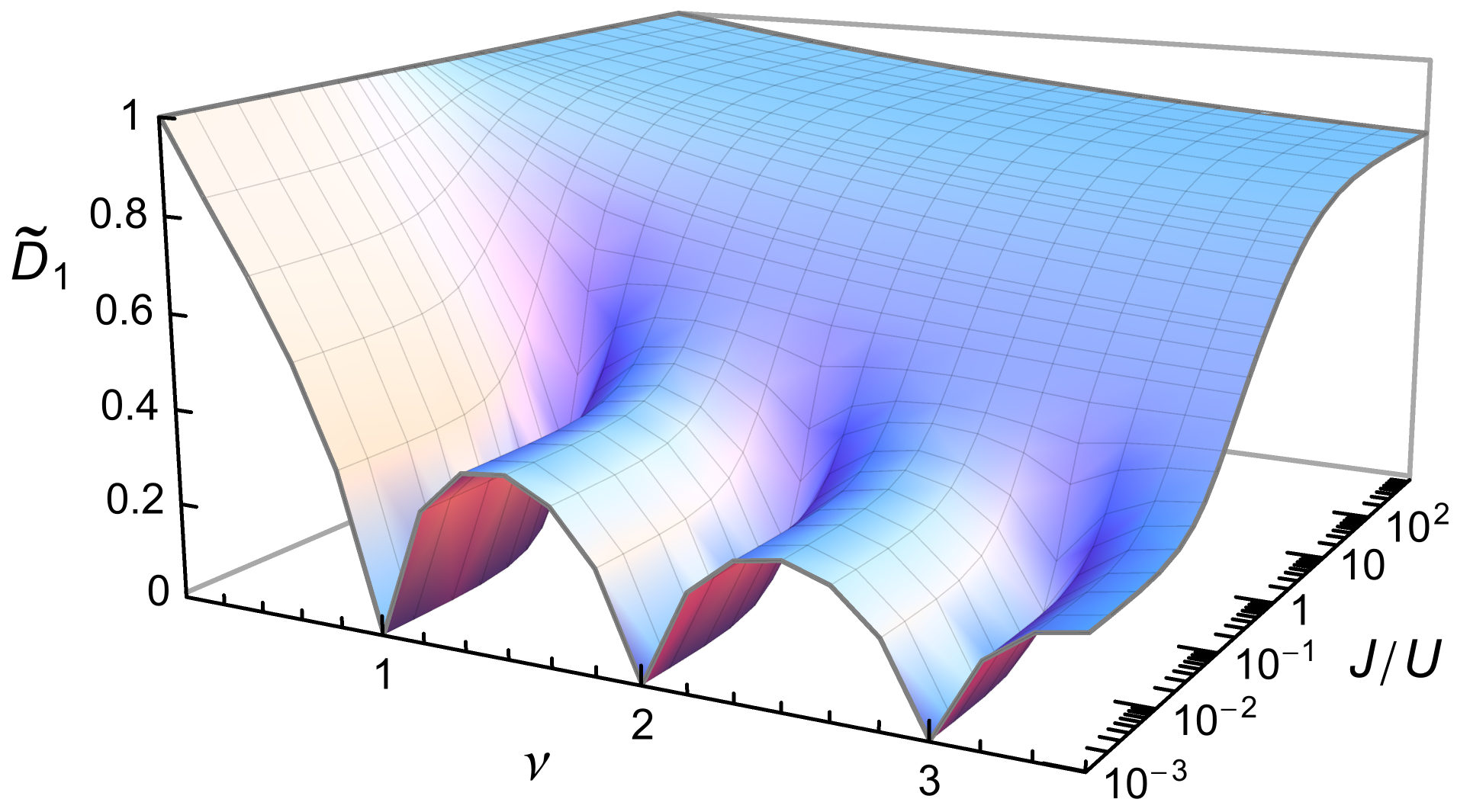}
 \caption{Finite-size generalized fractal dimension $\widetilde{D}_1$ for the ground state of the BHH with $L=6$ as function of $J/U$ and filling factor $1/6\leqslant \nu \leqslant 21/6$. The numerically calculated points correspond to the intersections of the grid lines shown on the surface.}
 \label{fig:D13D}
\end{figure}

In this work, all numerical results for $L\leqslant 10$ follow from ED. Figure \ref{fig:D13D} shows $\widetilde{D}_1$ for $L=6$ as a function of $\nu$ and $\eta$, and provides the three-dimensional perspective of Fig.~\ref{fig:DqDenplot} in the manuscript.
\begin{table}
 \caption{Information about the Pad\'e fits of the $\widetilde{D}_\infty$ and $\widetilde{D}_2$ data. The function \eqref{eq:pade} is determined by the expansion orders $(m,n)$ and has $m+n+1$ parameters, $N_D$ is the number of data points, $p$ is the goodness-of-fit, and $\eta_*$ the position of the maximum of the derivative of the fit with respect to $\eta$.}
 \begin{tabular}{cccccccccccc}
 \hline\hline\\[-4mm]  
 & & \hspace*{4mm} & \multicolumn{3}{c}{$\widetilde{D}_\infty$} & \hspace*{4mm} & \multicolumn{3}{c}{$\widetilde{D}_2$}  \\\hline 
 $L$ & $N_D$ & \hspace*{4mm} & $(m,n)$ & $p$ & $\eta_*$ & \hspace*{4mm} & $(m,n)$ & $p$ & $\eta_*$ \\\hline 
 $5$ & $55$ & & $(1,5)$ & -- & $0.16447$ & & $(1,5)$ & -- & $0.15714$ \\
 $6$ & $55$ & & $(3,3)$ & -- & $0.17339$ & & $(1,6)$ & -- & $0.16624$ \\
 $7$ & $55$ & & $(1,6)$ & -- & $0.17969$ & & $(3,4)$ & -- & $0.17334$ \\
 $8$ & $55$ & & $(1,6)$ & -- & $0.18513$ & & $(5,2)$ & -- & $0.17891$ \\
 $9$ & $55$ & & $(3,4)$ & -- & $0.18928$ & & $(3,5)$ & -- & $0.18328$ \\
 $10$ & $55$ & & $(5,2)$ & -- & $0.19274$ &  & $(3,5)$ & -- & $0.18709$ \\
 $12$ & $41$ & & $(3,2)$ & $0.35$ & $0.19967$ & & $(3,2)$ & $0.01$ & $0.19436$ \\
 $14$ & $41$ & & $(3,2)$ & $0.62$ & $0.20406$ & & $(3,3)$ & $0.17$ & $0.19835$ \\
 $16$ & $41$ & & $(3,2)$ & $0.08$ & $0.20777$ & & $(5,2)$ & $0.01$ & $0.20239$ \\
 $18$ & $41$ & & $(4,3)$ & $0.01$ & $0.21088$ & & $(5,2)$ & $0.05$ & $0.20563$ \\
 $20$ & $41$ & & $(3,3)$ & $0.02$ & $0.21376$ & &  & & \\
 $25$ & $36$ & & $(4,3)$ & $0.34$ & $0.21874$ & &  & & \\
 $30$ & $36$ & & $(4,3)$ & $0.04$ & $0.22256$ & &  & & \\
 \hline
 \end{tabular}
 \label{tab:Pfits}
\end{table}

As recently put forward \cite{Luitz2014,Luitz2014d,Luitz2014f,Luitz2014g}, for larger $L$, the moments $R_q$, and in turn the generalized fractal dimensions, can be obtained using quantum Monte Carlo (QMC). 
Choosing the Fock basis as computational basis, the intensity $|\psi_{\bm{n}}(\eta)|^2$ corresponds to the observational probability of the Fock state $\ket{\bm{n}}$ in the Monte Carlo sampling. Hence,  $\widetilde{D}_\infty$, which is entirely determined by the intensity of one known Fock state, can be efficiently estimated. 
Furthermore, for integer $q$, the moments can be evaluated using a replica trick: The sum $\sum_{\bm{n}}|\psi_{\bm{n}}(\eta)|^{2q}$ can be interpreted as the probability to observe the same Fock state at the same position in $q$ independent Markov chains. For example, after generating two Markov chains $\{\ket{\bm{n}_\alpha}\}$ and $\{\ket{\bm{m}_\alpha}\}$ of length $M$, the moment $R_2$ can be approximated by $R_2\simeq M^{-1}\sum_{\alpha=1}^M \delta_{\bm{n}_\alpha,\bm{m}_\alpha}$. 

The length of the Markov chains required for a desired relative uncertainty $\sigma_\text{rel}$ in the estimation of $\widetilde{D}_2$ can be determined by \cite{Lindinger2017c}
\begin{equation}
 M=\frac{\mathcal{N}^{\widetilde{D}_2}}{\left(\sigma_{\text{rel}} \widetilde{D}_2 \ln\mathcal{N}\right)^2}.
\end{equation}
The expression for $\widetilde{D}_\infty$ is formally the same, and we always set $\sigma_\text{rel}=10^{-3}$. The accessibility of the calculation depends on the expected values of the dimensions, very high values (corresponding to low observational probabilities) may require a too large $M$ making this approach inefficient. 
Since the $\widetilde{D}_q$ decrease with $q$, dimensions for larger $q$ are easier to calculate, implying that larger system sizes can be reached for $q=\infty$ than for $q=2$.
We use a QMC code based on the stochastic Green function algorithm \cite{Rousseau2008a,Rousseau2008}. 
Computing time depends on $\eta\equiv J/U$ and $L$, in particular the number of thermalizations required varies greatly with these parameters. In our simulations we set $\beta J=4L$ and check that thermalization has been achieved by analyzing the convergence and stability of the ground state energy. 
\begin{figure}
 \includegraphics[width=\columnwidth]{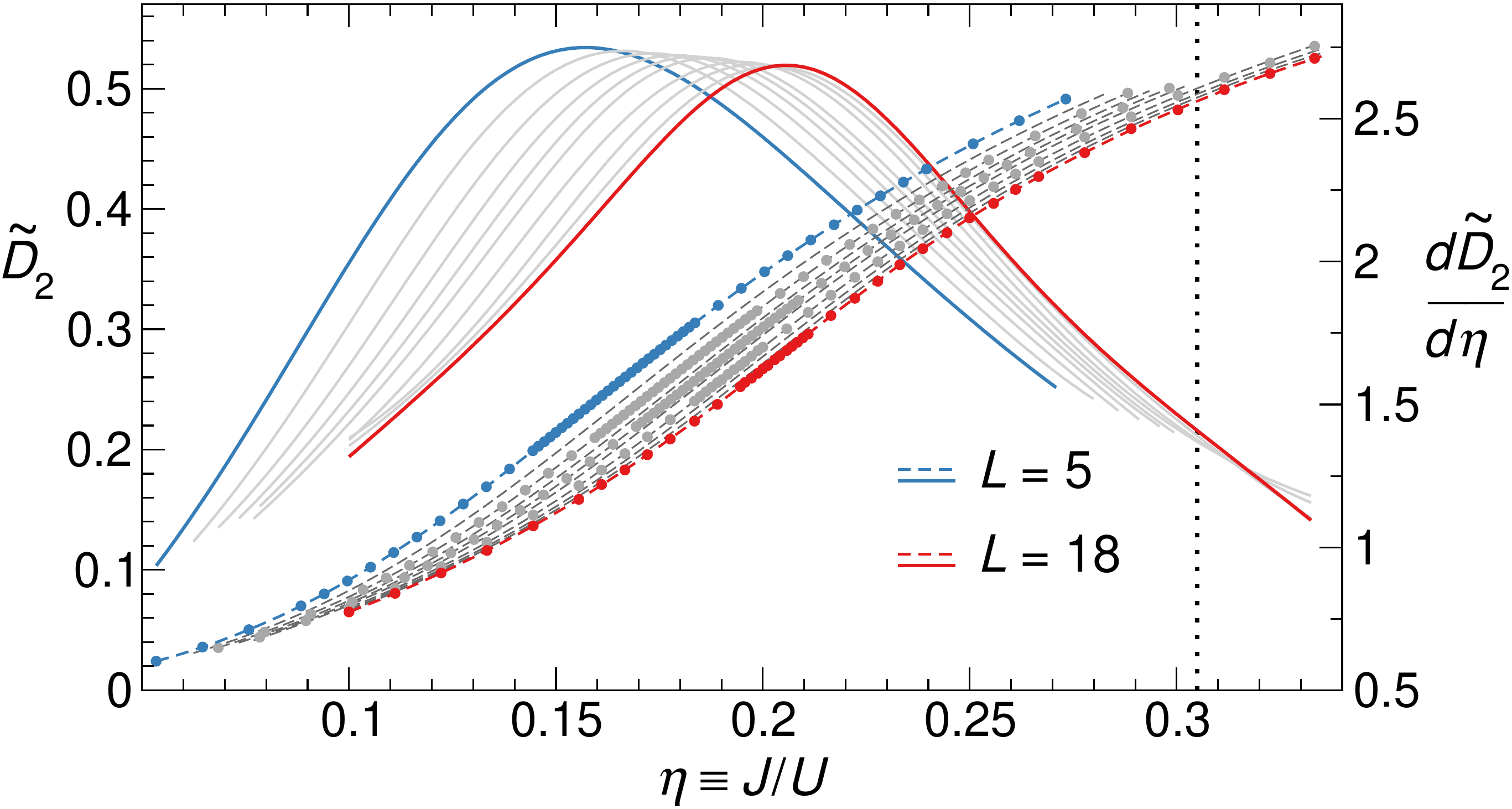}
 \caption{Dimension $\widetilde{D}_2$ (left vertical axis) and its derivative (right vertical axis) versus $\eta$ for the BHH ground state and $L=\{5\text{--}10,12,14,16,18\}$, $\nu=1$. Symbols indicate numerical data (errors within symbol size), dashed lines correspond to the best Pad\'e fits and solid lines mark their respective derivatives. For the sake of clarity, symbols are shown only for $L=\{5,7,9,12,18\}$. The vertical dotted line shows the position of the MI-SF transition in the thermodynamic limit \cite{Ejima2011}.}
 \label{fig:PadefitsD2}
\end{figure}

For the location of the maximum of the derivatives $\widetilde{D}'_q(\eta)\equiv d \widetilde{D}_q/d\eta$, we first find the best Pad\'e approximant 
\begin{equation}
 P_{m,n}(\eta)=\frac{\sum_{j=0}^m a_j \eta^j}{1+\sum_{j=1}^n b_j \eta^j}
 \label{eq:pade}
\end{equation}
that describes the $\widetilde{D}_q$ data. The fit is then differentiated and the maximum $\eta_*(L)$ numerically found. In order to increase the reliability of this procedure, we increase the number of $\widetilde{D}_q$ data points around the $\eta$-region where the maximum is expected by using a sampling step of $\Delta\eta=0.014$. In the fit we take into account the individual uncertainty of each data point, and we choose the simplest fit that provides a goodness-of-fit $p\geqslant 0.01$ \cite{Rodriguez2011}. We believe this to be a good criterion as we know that the uncertainties obtained from the QMC simulations are slightly underestimated. Data for $L\leqslant 10$, have the precision provided by ED, and we perform the fit without assuming errors in the data. In this case we take the first fit that yields $\chi^2<10^{-8}$. We check that all fits fulfilling this condition provide the same estimation for $\eta_*$. In Tab.~\ref{tab:Pfits}, we give the relevant data for the Pad\'e fits and the resulting value for $\eta_*(L)$. We set the uncertainty of $\eta_*(L)$ to be $\sigma=\Delta\eta/3$, i.e., solely determined by the sampling resolution in $\eta$ of the $\widetilde{D}_q$ data. The analyses of $\widetilde{D}'_\infty(\eta)$ and $\widetilde{D}'_2(\eta)$ are shown in Figs.~\ref{fig:PadefitsDinf} and \ref{fig:PadefitsD2}, respectively. All errors and confidence intervals provided in the manuscript follow from a bootstrap procedure \cite{Rodriguez2011}.
\bibliographystyle{apsrev4-1}
%
\end{document}